\documentclass[aps,pop,showpacs,twocolumn,superscriptaddress,letterpaper]{revtex4-1}
\usepackage{mathrsfs}
\usepackage{amsmath}
\usepackage{bm}
\usepackage{color}
\usepackage{graphicx}
\usepackage{hyperref}
\usepackage{float}
\usepackage{makecell}
\hypersetup{pdfpagemode=UseNone,pdfstartview=FitH,pdfstartpage=1}

\begin{document}


\title{ENN's Roadmap for Proton-Boron Fusion Based on Spherical Torus}
\author{
Min-sheng Liu*, Hua-sheng Xie**, Yu-min Wang, Jia-qi Dong, Kai-ming Feng, Xiang Gu, Xian-li Huang, Xin-chen Jiang, Ying-ying Li, Zhi Li, Bing Liu, Wen-jun Liu, Di Luo, Yueng-Kay Martin Peng, Yue-jiang Shi, Shao-dong Song, Xian-ming Song, Tian-tian Sun, Mu-zhi Tan, Xue-yun Wang, Yuan-ming Yang, Gang Yin and Han-yue Zhao
}
\thanks{{M. S. Liu (*liuminsheng@enn.cn) contributed to the commercialization roadmap, while H. S. Xie (**xiehuasheng@enn.cn) contributed to the scientific foundation. Both authors contributed equally to this work.}}

\affiliation{Hebei Key Laboratory of Compact Fusion, Langfang 065001, China}
\affiliation{ENN Science and Technology Development Co., Ltd., Langfang 065001, China}

\date{\today}

\begin{abstract}
ENN Science and Technology Development Co., Ltd. (ENN) is committed to generating fusion energy in an environmentally friendly and cost-effective manner, which requires abundant aneutronic fuel. Proton-boron ( p-$^{11}$B or p-B) fusion is considered an ideal choice for this purpose. Recent studies have suggested that p-B fusion, although challenging, is feasible based on new cross-section data, provided that a hot ion mode and high wall reflection can be achieved to reduce electron radiation loss. The high beta and good confinement of the spherical torus (ST) make it an ideal candidate for p-B fusion. By utilizing the new spherical torus energy confinement scaling law, a reactor with a major radius $R_0=4$ m, central magnetic field $B_0=6$ T, central temperature $T_{i0}=150$ keV, plasma current $I_p=30$ MA, and hot ion mode $T_i/T_e=4$ can yield p-B fusion with $Q>10$. A roadmap for p-B fusion has been developed, with the next-generation device named EHL-2. EHL stands for ENN He-Long, which literally means ``peaceful Chinese {Loong}". The main target parameters include $R_0\simeq1.05$ m, $A\simeq1.85$, $B_0\simeq3$ T, $T_{i0}\simeq30$ keV, $I_p\simeq3$ MA, and $T_i/T_e\geq2$. The existing ST device EXL-50 was simultaneously upgraded to provide experimental support for the new roadmap, involving the installation and upgrading of the central solenoid, vacuum chamber, and magnetic systems.  The construction of the upgraded ST fusion device, EXL-50U, was completed at the end of 2023, and it achieved its first plasma in January 2024. The construction of EHL-2 is estimated to be completed by 2026.
\end{abstract}


\maketitle

\section{Introduction}\label{sec:intro}

The quest for controlled fusion energy has been a longstanding dream for humanity since the 1950s, primarily due to its carbon-free nature, high unit energy, and abundance of resources. However, the challenges associated with achieving this goal are both scientific and engineering-related. Several books (cf. \cite{Chen2011}) and papers (cf. \cite{Wurzel2022}) review the progress in this field.

The development of a fusion energy reactor involves three essential steps: (1) selection of fusion fuels; (2) selection of a confinement approach; and (3) selection of a method to utilize or convert the energy produced. Over the past 70 years, much attention has been focused on the second step, and the feasibility or progress of a confinement approach can be described by the product of three physical parameters: ion density ($n_i$), ion temperature ($T_i$), and energy confinement time ($\tau_E$), commonly referred to as the Lawson criteria \cite{Lawson1957,Wurzel2022}. To confine plasmas efficiently, various approaches have been considered. Among these, magnetic confinement, which uses a magnetic field to confine the plasma, and inertial confinement, which initiates nuclear fusion reactions by compressing and heating fuel targets, are considered the two most promising techniques. Significant progress has been made in both magnetic confinement fusion and inertial fusion.
For example, in JET's 1997 Deuterium-Tritium (DT) experiments, a Q value, defined as the ratio between the output fusion power and the input power, of 0.67 was achieved \cite{Gibson1998, Reinders2021}. In the latest DT experiments in 2021, a value of 0.33 was achieved for 5 seconds, setting a new record for magnetic confined fusion energy output with 59 MJ \cite{Gibney2022}. In recent experiments at NIF, a Q value of approximately 1.5 was achieved with fusion energy output around 3.15 MJ \cite{Tollefson2022}, indicating that more fusion power was produced than input power, or ignition was achieved in inertial confinement fusion. { Though there are differences in the definition of ignition\cite{Wurzel2022} between the inertia confinement fusion and that of magnetic confinement fusion, there is little doubt about the scientific feasibility of D-T fusion.}

Considerable effort has been dedicated to magnetic confinement fusion. Various configurations have been proposed to confine plasmas, including tokamaks \cite{Wesson2004}, stellarators \cite{Nespoli2022}, field reversed configuration (FRC) \cite{Steinhauer2011}, and magnetic mirrors \cite{Post1987}. Tokamaks and stellarators are toroidal devices that utilize twisted magnetic fields to confine high-temperature plasma. Both configurations have demonstrated much higher plasma parameters than other magnetic confinement configurations to date. In stellarators, twisted magnetic field lines are created using external coils, while in tokamaks, field lines are twisted by a combination of the toroidal magnetic field generated by toroidal field coils (TF) and the poloidal magnetic field generated by poloidal field coils (PF) and the plasma current. Tokamaks are generally easier to construct than stellarators, and over one hundred tokamaks have been built and extensively studied both experimentally and theoretically. This has provided a solid foundation for the design of next-generation tokamaks such as ITER \cite{Ikeda2010} and CFETR \cite{Chan2015}.

In addition to conventional tokamaks with a large aspect ratio, defined as $A = R_0/a$ where $R_0$ and $a$ denote the major and minor radius, respectively, the spherical tokamak/torus (ST) with a small aspect ratio has been proposed \cite{Peng1986}. The small aspect ratio significantly influences magnetohydrodynamic (MHD) stabilities, allowing ST to operate with high $\beta_T$, the ratio of plasma pressure to toroidal magnetic pressure. This enables a more efficient use of the magnetic field, benefiting compact devices. Experiments have been conducted on START \cite{Gryaznevich1998}, MAST \cite{Sykes2001}, NSTX \cite{Menard2003}, and Globus-M \cite{Gusev2009}. Recent experimental results from the Globus-M2 device have demonstrated that energy confinement in ST can be significantly increased by doubling the toroidal magnetic field \cite{Kurskiev2022}. Furthermore, a new scaling law for ST has been developed, showing promising energy confinement time results at high toroidal magnetic fields, i.e., $\tau_E \propto I_p^{0.53} B_T^{1.05}$ \cite{Kurskiev2022}. One of the main drawbacks of ST is the limited space within the central solenoid (CS) coil, making it challenging to increase the toroidal field. However, recent advances in high-temperature superconductor (HTS) technology have made it possible to achieve high magnetic fields in compact devices {\cite{Creely2023,Hartwig2024}}. ST40, designed with a magnetic field of 3T, has operated at a toroidal field (TF) up to 2 T at $R_0$ = 40 cm, although the TF magnet is made from Cu and is liquid nitrogen (LN2) cooled \cite{Gryaznevich2022}. Recently, ion temperature plasmas of around  9.6 keV ($>100$ million degrees Kelvin) were achieved in ST40 \cite{McNamara2023}, representing a significant milestone for compact fusion devices.  The UK has also initiated the ambitious Spherical Tokamak for Energy Production (STEP) program (https://step.ukaea.uk/), which aims to develop a compact prototype reactor by 2040 capable of delivering a net electric power of $P_{el}>$100 MW to the grid\cite{Meyer2022}.

The p-$^{11}$B fusion has been studied both theoretically {\cite{Nevins1998, Geser2020,Putvinski2019,Ochs2024}} and experimentally{\cite{Margarone2022,Stave2011,Magee2023}}. On the experimental side, research has been conducted using laser-produced plasmas \cite{Margarone2022} and particle accelerators employing beam-target fusion \cite{Stave2011}. Recently, clear experimental measurements of $\alpha$ particles produced by p-$^{11}$B fusion in the Large Helical Device (LHD) have been achieved\cite{Magee2023}. This involved using high-energy neutral beams and boron powder injection in high-temperature fusion plasma. These findings represent the first experimental demonstration of p-$^{11}$B fusion in a magnetically confined plasma.

The significant accomplishments mentioned above have significantly enhanced confidence in the potential of commercial fusion energy. However, for fusion energy to become more competitive with other sources, such as fission and solar energy, considerations beyond the physics aspects are necessary. Achieving commercial fusion energy in the near future requires careful consideration of various options. In this work, we describe ENN's chosen roadmap for fusion energy, encompassing the selection of fuels, confinement approaches, and a research plan for the next 10-15 years.

This paper is organized as follows: In Sec. \ref{sec:fuel}, we discuss the choice of fuels. In Sec. \ref{sec:systemcode}, we describe a newly built system code that demonstrates the feasibility of a proton-Boron fusion energy reactor based on a spherical torus. In Sec. \ref{sec:tech}, we summarize the key technologies that need resolution. In Sec. \ref{sec:roadmap}, we present the ENN fusion roadmap. Finally, in Sec. \ref{sec:summ}, we provide a summary and discuss the findings.

\section{Choice of the Fuels}\label{sec:fuel}
In order to achieve commercial fusion energy, it is essential to utilize eco-friendly fusion sources. This implies a need for low neutron yield or aneutronic fusion. Additionally, the cost per unit of fusion power must be sufficiently low to compete with other energy sources such as wind, photovoltaic (PV), and fission. To fulfill these requirements, the fuel must be abundant and accessible, energy conversion must be high, and the technology must be suitable for the distributed energy market.

A comparison of the advantages and disadvantages of different fusion fuels is presented in Table \ref{tabel:comparison}. D-T fusion stands out as the easiest reaction to achieve fusion energy gain due to its high reaction cross section at relatively low temperatures, as well as its high energy yield for each reaction. Currently, the mainstream roadmap for commercial fusion is based on D-T fusion.

However, D-T fusion poses significant challenges, particularly concerning energetic neutron shielding and tritium breeding. Currently, there are no materials available for effective neutron shielding, and achieving a tritium breeding rate (TBR) larger than 1.2, which is necessary, may be difficult in the short term{\cite{Reinders2021}}. The proven reserves of  $^6$Li,  which is less than 10\% in natural lithium, for tritium breeding is also limited. Concerns about D-T fusion for energy have been raised from an engineering perspective \cite{Lidsky1983, Reinders2021}.  For example, fusion energy can only be extracted from the surface, which limits the power output per unit volume, and thus affects the unit cost. Even if neutron shielding and tritium breeding are well addressed, the two major drawbacks of fission—radioactivity safety or environment\cite{Reinders2021} concerns and limited resources—{hold similarly to D-T fusion, though at a much less severe level since fusion reactors are inherently safer than fission reactors, and lithium resources are abundant enough to support D-T fusion as a first-generation fusion fuel for about one thousand years.} These concerns, combined with the challenges mentioned above, make D-T fusion challenging to compete with  fission \cite{Lidsky1983} or fossil fuels\cite{Lerner2023}.

D-D fusion, on the other hand, is relatively easy to achieve due to its large reaction cross-section and abundance of fuel. However, it also produces high-energy neutrons. D-$^3$He fusion produces fewer neutrons and has a relatively large reaction cross-section, making it an attractive choice for fusion. Nevertheless, the disadvantages of D-$^3$He fusion include the side effects of D-D fusion and the scarcity of $^3$He storage on Earth. If $^3$He can be mined from the moon, it may make D-$^3$He fusion more feasible, although the reserves and cost of $^3$He remain open questions in the short term.

\begin{figure*}
	\centering
		\includegraphics[width = 0.99\textwidth]{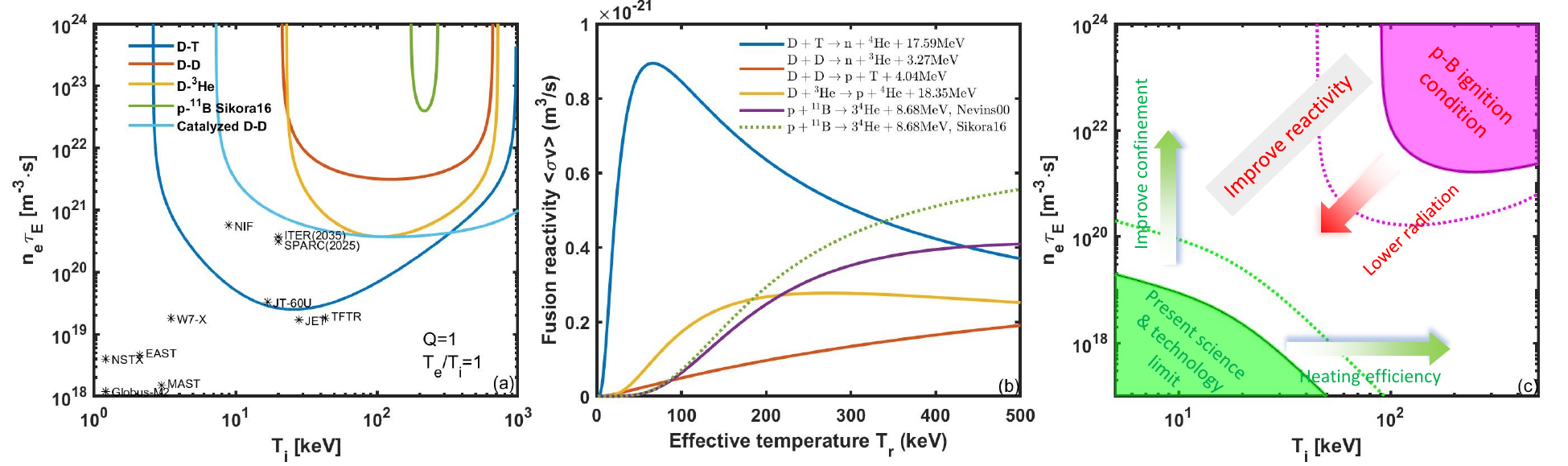}
	\caption{The triple products (a) and the thermal fusion reactivity (b) for D-T, D-D, D-$^3$He, and p-$^{11}$B fusion, and the strategy to make p-$^{11}$B fusion possible (c). {The data are from Refs. \cite{Wurzel2022, Xie2023, Xie2023a, Nevins2000, Sikora2016, Bosch1992}.}}
	\label{fig:tripleproduct}
\end{figure*}

The p-$^{11}$B fusion is an attractive fusion choice for several reasons. From the fuel perspective, abundant boron is readily available on Earth. Regarding fusion products, p-$^{11}$B fusion exclusively generates charged $\alpha$ particles, which have the potential for active control and direct electricity generation through charge separation. It's important to note that side reactions, such as p+$^{11}$B$\rightarrow$ n+$^{11}$C, producing low-energy neutrons with an energy of 2.765 MeV, exist. The fraction of this fusion reaction is approximately $10^{-5}$, and the low-energy neutrons are more manageable. These properties align with the requirements for commercial fusion. However, it should be acknowledged that the reaction threshold for p-$^{11}$B fusion is exceptionally high due to its relatively low fusion cross-section. The criteria for ignition and fusion reactivity for various fusion reactions are illustrated in Figure \ref{fig:tripleproduct}. Notably, the required criteria for p-$^{11}$B fusion are three orders of magnitude higher than those for D-T fusion. The fusion reactivity, $\langle\sigma v\rangle$, significantly improves when the effective temperature, {$T_{r} = (m_1T_2+m_2T_1)/(m_1+m_2)$}, exceeds 200 keV. Two sets of different cross-section data are utilized in p-$^{11}$B fusion reactivity calculations {\cite{Nevins2000, Sikora2016}}. Despite these challenges, { it is believed that high ion temperatures or energies can be achieved by utilizing high-energy negative ion-based neutral beam injection (NNBI) \cite{Ikeda2006} and effective ion cyclotron range of frequency (ICRF) heating\cite{Kirov2024}.} Notably, both TFTR \cite{Hawryluk1998} and JT-60 \cite{Ishida1996} tokamaks have achieved ion temperatures $T_i \simeq 45$ keV.

After this analysis, choosing p-$^{11}$B as the fuel for commercial fusion appears reasonable. The primary reason is not that it is easy, but rather that we have limited choices. Due to the scarcity of fuels, options like D-T and D-$^3$He are not viable. Additionally, to avoid dealing with neutrons, D-D fusion is not a preferred choice.

\begin{table*}
{\small
	\begin{center}
	\caption{The comparison between different fusion fuels. {Note that accurate fuel prices and reserves are difficult to obtain and may change over time; therefore, the data provided here are rough estimates for reference purposes. {See Appendix \ref{sec:fuelcost} for more details.}}}
	\begin{tabular}{c|c|c|c|c|c}
	\hline
	\textbf{ \makecell{Nuclear\\ reaction}} & \textbf{Advantages} &  \textbf{Disadvatages} & \textbf{\makecell{Rough fuel price \\({CNY/g})}} &\textbf{\makecell{Output energy \\ {20TW$\cdot$year}}} &\textbf{Rough reserves}\\
	\hline
	D-T & Easiest way to fusion & \makecell{ High energy neutron\\ wall material protection,\\ Tritium breeding}
	& T: {{0.2} million} & \makecell{{T: 1000 tons} \\{$^6$Li: 2000 tons}}& \makecell{{T: 25 kg}\\{$^6$Li: 2 million tons} }\\
	\hline
	D-D & Relatively easy & \makecell{ High energy neutrons,\\ peak  reaction cross \\section low }
	& D: 30 &{ D: 7000 tons} & 45 trillion tons\\
	\hline
	D-$^3$He & \makecell{few neutrons, \\low neutron energy} & \makecell{$^3$He expensive,\\D-D side effects }
	&$^3$He: {{0.1} million} & { $^3$He: 1000 tons} & \makecell{produced by\\ nuclear reactions,\\ mined on the moon}\\
	\hline
	p-$^{11}$B & \makecell{With few neutrons \\ and cheap raw fuels,\\ $\alpha$ particles \\direct energy coversion} & \makecell{Extreamely high\\ reaction threshold}
	&$^{11}$B: {30} & {$^{11}$B: 8000 tons} & \makecell{{1 billion tons}}\\
	\hline
	\end{tabular}
	\label{tabel:comparison}
	\end{center}}
\end{table*}

\section{System code calculations}\label{sec:systemcode}
  
System codes utilizing scaling laws from experiments have been developed and employed \cite{Shimada2007, Costley2015}. However, to design a spherical tokamak using p-$^{11}$B as fuel, a new system code is required. This section will introduce our new system code based on the updated spherical tokamak scaling.

\subsection{Flow chart of the code}

\begin{figure*}
\centering
\includegraphics[width = 1.0\textwidth]{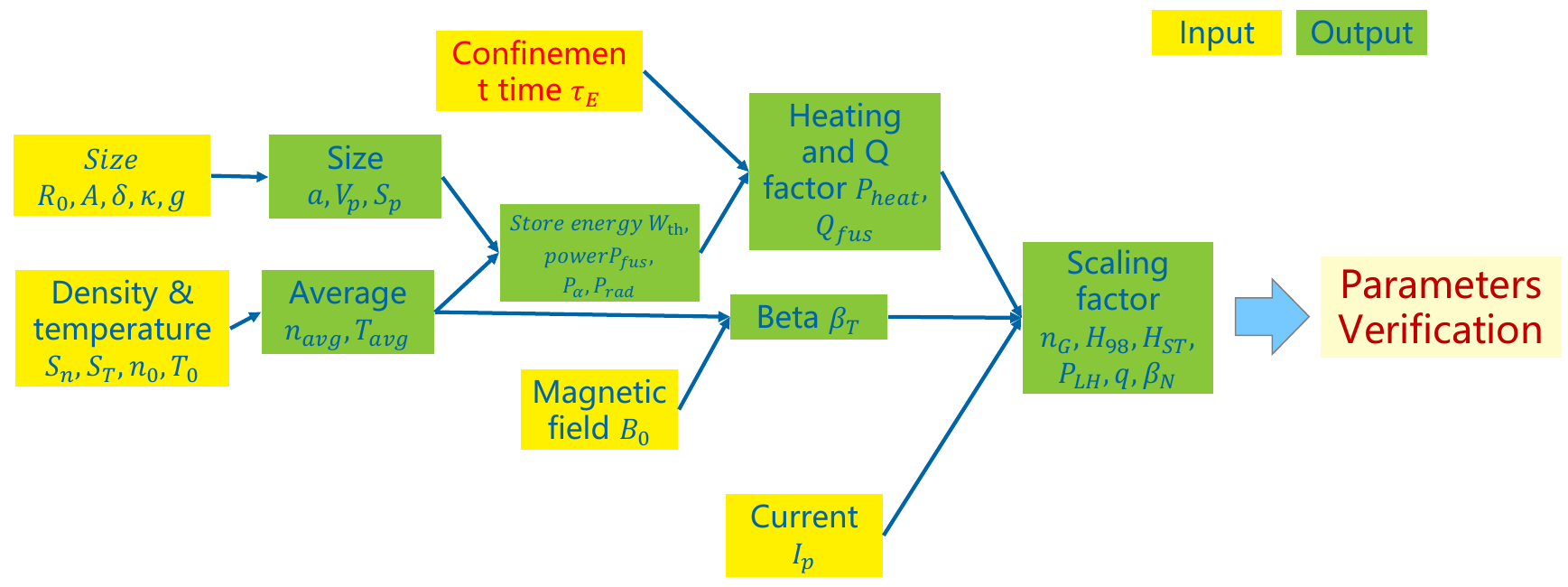}
\caption{The flow chart of the system code. A major difference from other system codes is that certain scaling law-relevant parameters, such as $\tau_E$, are set as input parameters to avoid sensitivity. }\label{fig:flowchart}
\end{figure*}

The new code is developed and streamlined from our original system code \cite{Cai2022}. The flow chart of the code is illustrated in Figure \ref{fig:flowchart}.

The input parameters encompass the size of the vacuum vessel or the shape of the plasma (explained in detail in subsection \ref{sec:shape}), density and temperature profiles, toroidal magnetic field ($B_0$), plasma current ($I_p$), and energy confinement time ($\tau_E$). The profiles are expressed as
\begin{eqnarray}
n(x) = n_0 (1-x^2)^{S_n},\\
T(x) = T_0 (1-x^2)^{S_T},
\end{eqnarray}
where $n_0$ and $T_0$ are the density and temperature on the magnetic axis, and $S_n$ and $S_T$ are the shape factors of density and temperature. Here, $x=r/a$ represents the normalized radial displacement, where $r$ is the displacement from the magnetic axis along the radial direction, and $a$ is the minor radius. The profiles can exhibit more H-mode-like behavior when decreasing the shape factors, i.e., having a steep gradient in the edge region. By using these expressions, the volume-averaged and line-averaged plasma density and temperature can be calculated as follows:
\begin{eqnarray}
\langle n\rangle = \int_0^1n_0\left(1-x^2\right)^{S_n}2x dx = \frac{n_0}{1+S_n},\\
\langle T\rangle = \int_0^1T_0\left(1-x^2\right)^{S_T}2x dx = \frac{T_0}{1+S_T},\\
\langle n\rangle_l = \int_0^1n_0\left(1-x^2\right)^{S_n}dx =\frac{\sqrt{\pi}}{2}\frac{\Gamma(S_n+1)}{\Gamma(S_n+1.5)}n_0,\\
\langle T\rangle_l = \int_0^1T_0\left(1-x^2\right)^{S_T}dx =\frac{\sqrt{\pi}}{2}\frac{\Gamma(S_T+1)}{\Gamma(S_T+1.5)}T_0.
\end{eqnarray}
{ Here, $\Gamma(z)$ is Euler's Gamma function.}

The energy confinement time is also an input parameter in the new code, enabling the calculation of heating power and fusion power.

The output parameters of the new code include plasma size, averaged density and temperature ($n_{avg}$ and $T_{avg}$), stored energy ($W_{th}$), fusion, $\alpha$ particle power, and radiation power ($P_{fus}$, $P_{\alpha}$, and $P_{rad}$). Combining the fusion power and energy confinement time allows the determination of the required heating power ($P_{heat}$) from the power balance equation, and the fusion energy gain factor ($Q_{fus}$) can be calculated. The plasma $\beta_T$ can be determined using profiles and toroidal magnetic field. Finally, scaling factors such as Greenwald density fraction ($n_G$), H factors ($H_{98}$ and $H_{ST}$), L-H mode transition power ($P_{LH}$), safety factor ($q$), and normalized beta ($\beta_N$) can be computed.

\subsection{Plasma shape}
\label{sec:shape}

\begin{figure}
\centering
\includegraphics[width=9cm]{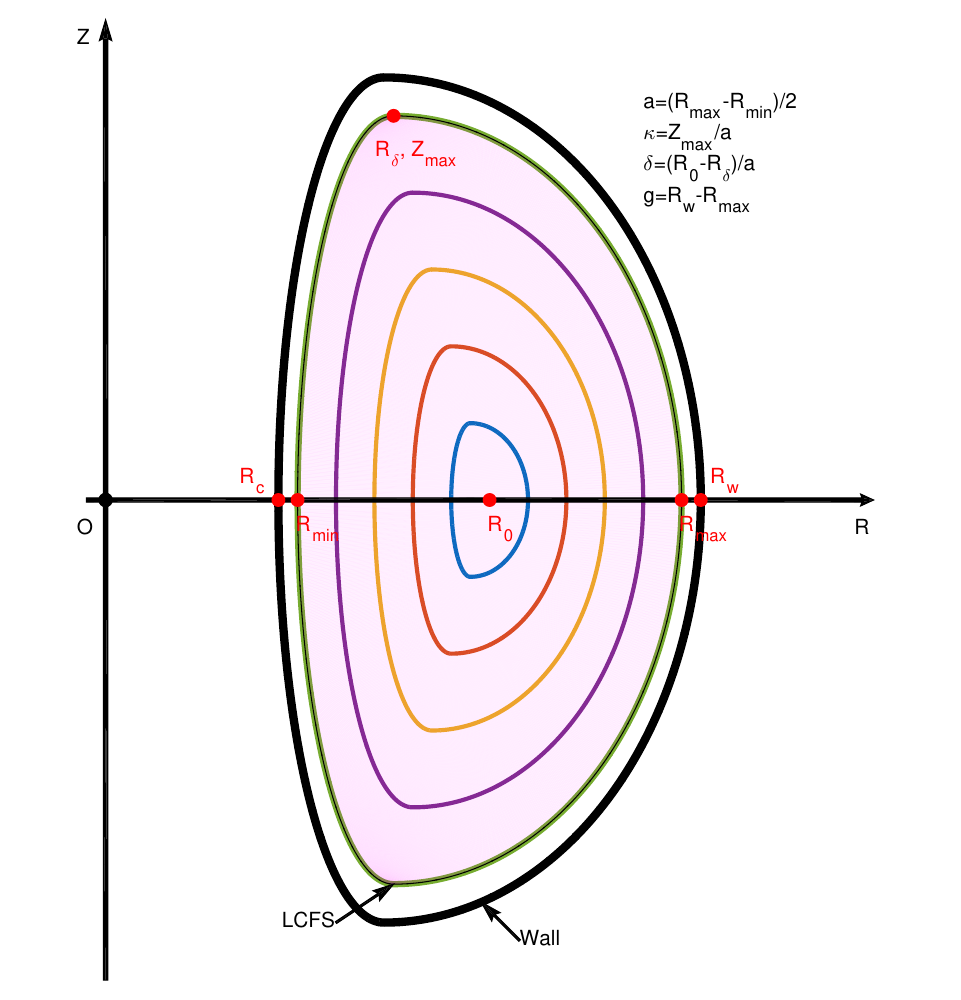}
\caption{The plasma shape.}\label{fig:shape}
\end{figure}

The geometry of the plasmas is illustrated in Figure \ref{fig:shape}. The shape of the plasma can be fixed when parameters such as the major radius ($R_0$), aspect ratio ($A{=R_0/a}$), triangularity ($\delta$), elongation ($\kappa$), and gap between the plasma and the vacuum wall ($g$) are specified. The initial design of the vacuum vessel can be determined based on these parameters. The plasma volume can be calculated using
\begin{eqnarray}
    V_p = 2\pi^2\kappa\left(R_0-\delta a\right)a^2+\frac{16}{3}\delta\pi \kappa a^3.
\end{eqnarray}
The volume of the device is calculated by using 
\begin{eqnarray}
    V_d = 2\pi^2\kappa\left(R_0-\delta\left( a+g\right)\right)\left( a+g\right)^2 +\frac{16}{3}\delta\pi \kappa (a+g)^3.
\end{eqnarray}

The surface area of the plasma, $S_p$, and the surface area of the device, $S_w$, are
\begin{eqnarray}
    S_p = \left(4\pi^2\frac{R_0}{a}\kappa^{0.65}-4\kappa\delta\right)a^2,
\end{eqnarray}
\begin{eqnarray}
    S_w = \left(4\pi^2\frac{R_0}{a+g}\kappa^{0.65}-4\kappa\delta\right)(a+g)^2.
\end{eqnarray}

It should be noted that we have set the major radius and the magnetic axis at the center of the plasma, i.e., $R_0=(R_{max}+R_{min})/2$. In realistic plasma equilibrium, the magnetic axis would experience a shift outside due to the Shafranov shift. We omit this difference for simplification.

\subsection{Species}
There are multiple species inside the plasma, including two ions for fusion reactions, helium particles, and other impurities, with their densities represented by $n_1$, $n_2$, $n_{He}$, and $n_{imp}$, respectively, and the charge numbers represented by $Z_1$, $Z_2$, $Z_{He}$, and $Z_{imp}$. The ion and electron density can be determined according to the quasi-neutral condition, i.e.,
\begin{eqnarray}
n_i=\frac{n_1+n_2}{1+\delta_{12}}+n_{He}+n_{imp},\\
n_e = \frac{n_1Z_1+n_2Z_2}{1+\delta_{12}}+n_{He}Z_{He}+n_{imp}Z_{imp},
\end{eqnarray}
where $\delta_{12}=1$ when fusion particles 1 and 2 are identical, while $\delta_{12}=0$ when they are different. The average and effective charge numbers can be calculated by
\begin{eqnarray}
Z_i=\frac{n_e}{n_i},\\
Z_{eff} = \frac{\frac{n_1Z_1^2+n_2Z_2^2}{1+\delta_{12}}+n_{He}Z_{He}^2+n_{imp}Z_{imp}^2}{n_e}.
\end{eqnarray}

The effective density of the two fusion ions is presented by
\begin{equation}
n_{12} = \frac{n_1+n_2}{1+\delta_{12}},
\end{equation}
and the ratio of the fusion ions is $f_{12} = n_{12}/n_i$. The ratios of the helium ions and impurities are represented by $f_{He} = n_{He}/n_i$ and $f_{imp} = n_{imp}/n_i$, respectively. Furthermore, the ratios of the two separate fusion ions can be calculated by $x_1 = n_1/n_{12}$ and $x_2 = n_2/n_{12}$. Under this definition, and together with plasma quasi-neutrality, the following relations need to be satisfied,
\begin{eqnarray}
f_{12}+f_{He}+f_{imp} =1,\\
\frac{f_{12}(x_1Z_1+x_2Z_2)}{1+\delta_{12}}+f_{He}Z_{He}+f_{imp}Z_{imp}=1.
\end{eqnarray}
Assuming that the temperature of the ions is the same, while the temperature ratio of the electrons and ions is fixed at $f_T$, i.e., $T_e = f_T T_i$, the effect of the ion and electron temperature ratio can be evaluated.

\subsection{Fusion power}
 
The fusion power of the two fusion ions can be calculated by
\begin{equation}
P_{fus} = \frac{1}{1+\delta_{12}} n_1 n_2 \langle \sigma v\rangle Y = \frac{Y}{1+\delta_{12}}n_{10}n_{20}\Phi V_p,
\end{equation}
where $Y$ is the total energy released from the fusion reaction, and $\sigma$ represents the cross-section of the two fusion ions. Here $n_{10}$ and $n_{20}$ are the densities at the magnetic axis for the two fusion ions. Considering the radial distribution of the ion temperature, the fusion reactivity can be calculated as follows,
\begin{equation}
\Phi = 2\int_0^1(1-x^2)^{2S_n}\langle\sigma v\rangle xdx.
\end{equation}
where $\langle\sigma v\rangle$ is related to the effective ion temperature $T_i(x)$. Assuming there is a potential enhancement factor $f_\sigma$ \cite{Xie2023}, for non-Maxwellian distributed ions or from other effects, the fusion rate can be presented by 
\begin{equation}
\langle\sigma v\rangle = f_\sigma\langle\sigma v\rangle_M,
\end{equation}
where $\langle\sigma v\rangle_M$ can be used to calculate the fusion rate for a Maxwellian ion temperature distribution. In this way, the effect of the fusion rate enhancement can be evaluated.


The fusion equation for D-T fusion is
\begin{equation}
{\rm D+T}\rightarrow ^4_2{\rm He (3.5MeV)} + ^1_0n {\rm (14.1 MeV)}.
\end{equation}
Here $Y = 17.6$ MeV, and the formulas for D-T fusion cross-section are used \cite{Bosch1992}.


The fusion equation for p-$^{11}B$ fusion is
\begin{equation}
{\rm p+}^{11}{\rm B}\rightarrow 3 ^4_2{\rm He (8.7MeV)}.
\end{equation}
Here $Y = 8.7$ MeV, {and we use the new fusion cross-section data \cite{Sikora2016} instead of the old one \cite{Nevins2000}.}

After calculating the fusion power, the fusion gain factor can be obtained by
\begin{equation}
Q = \frac{P_{fus}}{P_{aux}},
\end{equation}
where $P_{aux}$ is the auxiliary power used to sustain the plasma.

\subsection{Power balance}
The power balance equation is given as 
\begin{equation}
\frac{dW_{th}}{dt} = P_{\alpha}+P_{aux}- P_{brem}-P_{cycl}-\frac{W_{th}}{\tau_E}=0,
\end{equation}
where $W_{th}$ is the plasma thermal energy, and $P_\alpha$ is the heating power by $\alpha$ particles.
$P_{brem}$ and $P_{cycl}$ are the bremsstrahlung and cyclotron radiation loss power, respectively.


In the non-relativistic case, the bremsstrahlung power can be evaluated using the following expression \cite{Nevins1998}
\begin{eqnarray}
&&P_{brem} \\\nonumber
&&= C_B n_{e0}^2\sqrt{k_BT_{e0}}V_p\left\{Z_{eff}\left[\frac{1}{1+2S_n+0.5S_T}+ \right.\right.\nonumber\\
&& \frac{0.7936}{1+2S_n+1.5S_T}\frac{k_BT_{e0}}{m_ec^2} + \left.\frac{1.874}{1+2S_n+2.5S_T}\left(\frac{k_BT_{e0}}{m_ec^2}\right)^2\right]\nonumber \\
&&+\left.\frac{3}{\sqrt{2}(1+2S_n+1.5S_T)}\frac{k_BT_{e0}}{m_ec^2}\right\}(MW),
\end{eqnarray}
where $C_B = 5.34\times10^{-43}$, $k_BT_{e0}$ and $m_ec^2$ are in keV, and the electron density in $m^{-3}$. Here $k_B$ and $m_e$ are the Boltzmann constant and electron rest mass, respectively.

The cyclotron power can be obtained from \cite{Kukushikin2009}
\begin{eqnarray}
P_{cycl}&=&4.14\times10^{-7}B_{T0}^{2.5}(1-R_w)^{0.5}V_pa_{eff}^{-0.5}\times\nonumber\\
&&n_{eff}^{0.5}T_{eff}^{2.5}\left(1+\frac{2.5T_{eff}}{511}\right)(MW).
\end{eqnarray}
Here, $R_w$ denotes the reflection rate of the wall, and it is important to reuse the cyclotron power by reflecting it back to the plasma. The effective density, $n_{eff}$, the effective minor radius, $a_{eff}$, and the effective temperature, $T_{eff}$, are defined as
\begin{eqnarray}
n_{eff} = \langle n_e\rangle = \frac{n_{e0}}{1+S_n},\\
a_{eff} = a\kappa^{0.5},
T_{eff} = T_{e0}\int_0^1(1-x^2)^{S_T}dx.
\end{eqnarray}
Note that the definition of $T_{eff}$ is different from $\langle T_e\rangle$.

The energy confinement time, $\tau_E$, is a critical parameter in magnetic confinement fusion and serves as an input parameter in the new system code. The scaling law for the { H-mode} energy confinement time in conventional tokamaks is described by the IPB98(y,2) model \cite{Shimada2007}, expressed as
\begin{equation}
\tau_{E}^{CT} = 0.0562 I_p^{0.93}B_T^{0.15}P_{loss}^{-0.69}n_e^{0.41}M^{0.19}R_0^{1.97}\epsilon^{0.58}\kappa^{0.78}
\end{equation} 
where $M$ is the average mass number { and $\epsilon=1/A$}. 

For spherical tokamaks, the energy confinement time can be described by the following scaling law \cite{Kurskiev2022}
\begin{equation}\label{eq:taust}
\tau_E^{ST} = 0.066I_p^{0.53}B_T^{1.05}P_{loss}^{-0.58}n_e^{0.65}R_0^{2.66}\kappa^{0.78}.
\end{equation}
Both of these scaling laws are used in the new system code to compare the different confinement factors between conventional tokamaks and spherical tokamaks. The H-factor, defined as the ratio of the required energy confinement time to the scaling, is utilized to evaluate the confinement of the tokamak. It can be calculated as follows,
\begin{eqnarray}
H_{98} = \frac{\tau_E}{\tau_E^{CT}},\\
H_{ST} = \frac{\tau_E}{\tau_E^{ST}}.
\end{eqnarray}
{ Note that there is not much difference between the L-mode and H-mode for ST scaling law; that is, Eq. (\ref{eq:taust}) includes the data for both.}

\subsection{Output parameters}
The toroidal beta can be calculated using
\begin{equation}
\beta_T = \frac{2\mu_0k_B\int(n_iT_i+n_eT_e)dV}{B_{T}^2}=\frac{2\mu_0k_B(n_iT_i+n_eT_e)}{(1+S_n+S_T)B_{T}^2},
\end{equation}
where $\mu_0$ is the permeability of free space.

The safety factor is calculated by default using \cite{Costley2015}
\begin{equation}
q = \frac{2\pi\kappa B_Ta^2\kappa}{\mu_0R_0I_p}= \frac{5B_Ta^2\kappa}{R_0I_p},
\end{equation}
and also with ST correction
\begin{equation}
q^* = \frac{\pi (1+\kappa^2)B_Ta^2\kappa}{\mu_0R_0I_p}.
\end{equation}
Generally, the safety factor should be larger than 2 to avoid disruption. The normalized beta of the plasma can be calculated using
\begin{equation}
\beta_N = 100\beta_T\frac{aB_T}{I_p},
\end{equation}
and it is required that $\beta_N < 12/A$ to stabilize instabilities.

\subsection{Code validation}

The ITER design parameters are employed to benchmark the new system code, and the comparison results are presented in Table \ref{tabel:CodeBenchmak}. The outcomes obtained from the new simplified system code closely align with the results reported by Costley \cite{Costley2015} for the same input parameters.

\begin{table}
	\begin{center}
	\caption{Comparison of the new system code with Costley's results for ITER parameters.}
	\begin{tabular}{|c|c|c|}
	\hline
	\textbf{ ITER parameters} & \textbf{Costley15} &  \textbf{ENN model} \\
	\hline
	\makecell{Central ion \\temperature, $T_{i0}$ (keV)} & 25 & 25\\
	\hline
	Central density, $n_{e0}$ (m$^{-3}$) & $0.77\times 10^{20}$ & $0.77\times 10^{20}$\\
	\hline
	Confinement time, $\tau_E$ (s) & 2.0 & 2.0\\
	\hline
	Beta, $\beta$ & 0.02 & 0.02\\
	\hline
	Magnetic field, $B_0$ (T) & 5.18 & 5.18\\
	\hline
	Major radius, $R_0$ (m) & 6.35 & 6.35\\
	\hline
	Aspect ratio, $A$ & 3.43 & 3.43\\
	\hline
	\makecell{Plasma currrent,\\ $I_p$  (MA)} & 9.2 & 9.2\\
	\hline
	\makecell{Heating power, \\$P_{heat}$ (MW)} & 70 &\textbf{ 70.4}\\
	\hline
	\makecell{Fusion power, \\$P_{fus}$ (MW)} & 350 & \textbf{356}\\
	\hline
	Fusion gain, $Q$ & 5 & \textbf{5.1}\\
	\hline
	\end{tabular}
	\label{tabel:CodeBenchmak}
	\end{center}
\end{table}

\section{Technologies to be resolved}\label{sec:tech}

The general strategy to make p-$^{11}$B fusion possible is shown in Fig. \ref{fig:tripleproduct}c, i.e., reduce the requirement by increasing the reactivity and reducing the radiation, and increasing the triple products. In pursuit of p-$^{11}$B fusion, it is imperative to enhance the triple product, particularly by elevating the ion temperature and lowering the reaction threshold conditions. While high $T_i$ is essential for increased fusion reactivity, it is crucial to manage electron bremsstrahlung loss power, which can surpass fusion power if $T_e$ matches $T_i$ . Hence, hot ion modes with $T_i/T_e=4$ are deemed necessary \cite{Cai2022}. { It is encouraging to note that in p-$^{11}$B fusion, the fusion product $\alpha$ particles primarily heat ions rather than electrons\cite{Dawson1981,Xie2023a}, contrasting with the dynamics in D-T fusion. This phenomenon can be attributed to the significantly closer velocities between the MeV $\alpha$ particles and the hundred keV ions compared to electrons, facilitating more efficient energy exchange, specifically heating, between ions and $\alpha$ particles.}

\begin{figure*}
\centering
\includegraphics[width=0.9\textwidth]{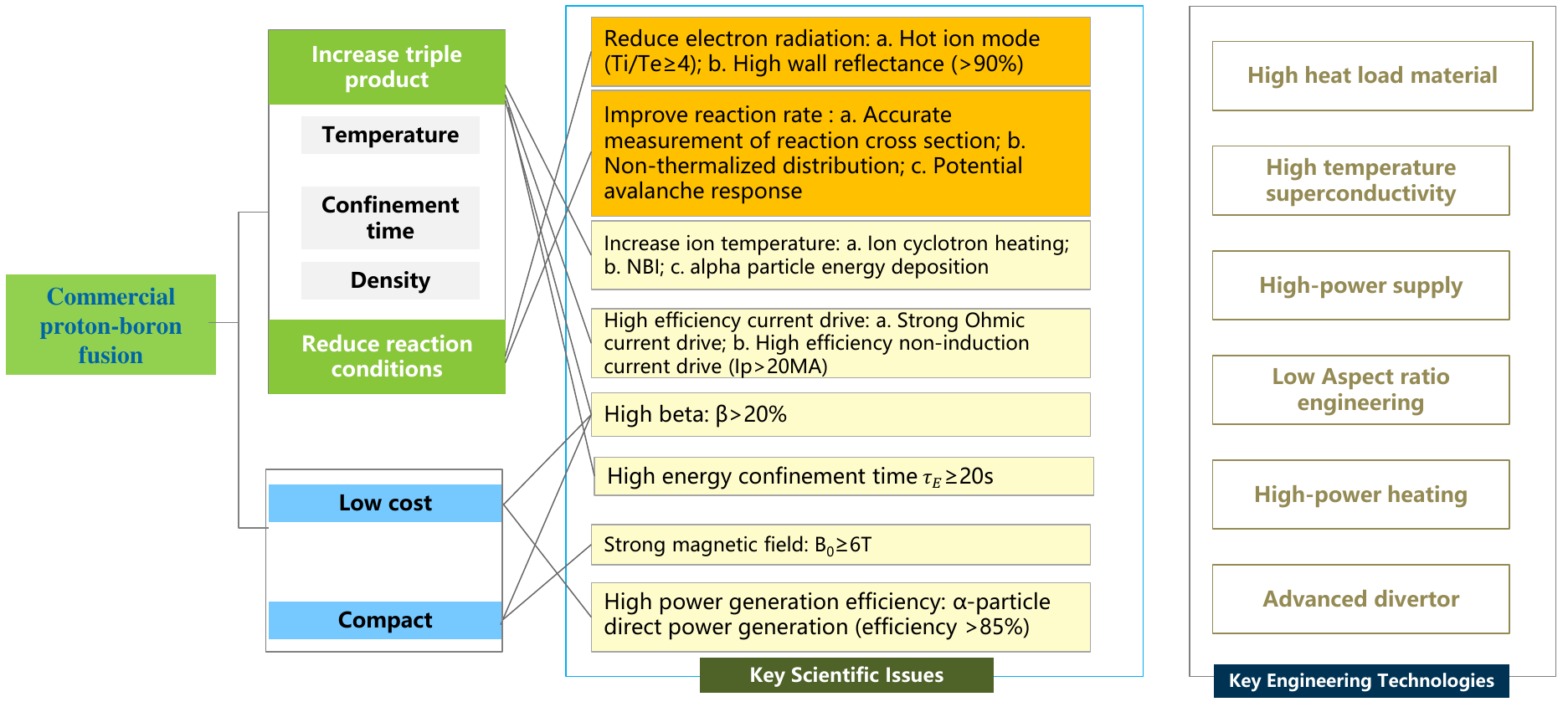}
\caption{Key issues for spherical torus p-B fusion. Primary factors can interact with each other. If one technological breakthrough is achieved, other technical requirements may be lowered accordingly.}\label{fig:SCiReq}
\end{figure*}

Augmenting the fusion reactivity becomes feasible with a larger fusion reaction cross-section{\cite{Sikora2016,Putvinski2019}}. Additional physical mechanisms, such as nonthermalized distribution \cite{Xie2023, Kong2024,Xie2024} and potential avalanche processes \cite{Hora2017} { and other nonlinear effects\cite{Wei2023}} in p-$^{11}$B fusion reactions, hold promise for further boosting fusion reactivity. { Recent JET D-T experiments have shown promising results. These include observations of synergistic effects on fusion rates with both NBI and ICRF \cite{Kirov2024}, as well as significant non-thermal beam-target and beam-beam fusion reaction\cite{Kiptily2023}.} The primary scientific requirements for realizing p-$^{11}$B fusion are succinctly outlined in Figure \ref{fig:SCiReq}.

Using the system code described in Section \ref{sec:systemcode}, the optimal parameters for achieving p-$^{11}$B fusion gain ($Q>1$) are as follows: magnetic field strength ($B_0$) in the range of $4-7$ T, ion temperature to electron temperature ratio ($T_i/T_e$) of $\geq3$, ion temperature ($T_{i0}$) between $100-300$ keV, electron density ($n_{e0}$) of $1-3\times10^{20}$ m$^{-3}$, and energy confinement time ($\tau_E$) ranging from $20-50$ s. Additionally, if the potential fusion enhancement factor $f_{\sigma}$ is high (e.g., $f_{\sigma}>2$), the other requirements can be significantly reduced. By utilizing the new ST energy confinement scaling law, a reactor with major radius ($R_0$) of $4$ m, central magnetic field ($B_0$) of $6$ T, central temperature ($T_{i0}$) of $150$ keV, plasma current ($I_p$) of $30$ MA, and hot ion mode ($T_i/T_e$) of $4$ can yield p-$^{11}$B fusion with $Q>10$. More detailed discussions of p-B reactor parameters can be found in Appendix \ref{sec:para}. {One should be careful that different assumptions may yield different results.}

The attainment of an extremely high p-$^{11}$B fusion gain is challenging. To illustrate the difficulty, we compare the cases of Conventional Tokamaks (CT) and {ST} in Table \ref{tabel:CTSTcmp} with a set of highly optimized conditions for p-$^{11}$B fusion, specifically $f_{\sigma}=5$ and $T_i/T_e=4$. The primary distinctions between CT and ST lie in the confinement scaling law and plasma beta. As a result, we observe that CT can only achieve $Q\simeq3$ with $B_0=12$ T, necessitating an even higher H factor of $H_{98}$=3.41. In contrast, ST can achieve $Q=30$ with $H_{ST}\simeq1$.

Benefiting from the high $\beta_T$ and favorable energy confinement time of ST, the required magnetic field and heating power are significantly lower compared to conventional tokamaks. Consequently, spherical tokamaks emerge as more suitable candidates for p-$^{11}$B fusion. However, it's crucial to note that currently, the plasma current, temperature, and energy confinement time of spherical tokamaks are considerably lower than those of conventional tokamaks and need validation at high parameters. Verifying the confinement properties at elevated plasma parameters constitutes a crucial task in ENN's fusion roadmap. Additionally, it is emphasized that other magnetic confinement approaches typically exhibit even lower confinement time scaling than tokamaks and, therefore, may not be a prioritized option in ENN's strategy.

\begin{table}
	\begin{center}
	\caption{The comparison of parameters of a CT and ST from the system code.}
	\begin{tabular}{|c|c|c|}
	\hline
	\textbf{ Parameters} & \textbf{CT} &  \textbf{ST} \\
	\hline
	\makecell{Average ion \\temperature, $T_{i,avg}$(keV)} & 33 & 33\\
	\hline
	\makecell{Average electron density,\\ $n_{e,avg} (10^{20}m^{-3})$} & 1.65  &0.66\\
	\hline
	Confinement time, $\tau_E(s)$ & 5 & 19.3\\
	\hline
	Beta, $\beta_T$ & 0.038 & 0.32\\
	\hline
	\makecell{Central Magnetic field,\\ $B_0(T)$} & 12 & 2.6\\
	\hline
	Major radius, $R_0(m)$ & 4 & 3.2\\
	\hline
	Fusion Gain, $Q$ & 3.2 & 30\\
	\hline
	Aspect ratio, $A$ &3.5 & 1.7\\
	\hline
	$n_0/n_G$ & 1 & 0.76 \\
	\hline
	Safty factor, q & 3.27& 2.16\\
	\hline
	Elongation, $\kappa$ &2.5 &3.3 \\
	\hline
	Fusion power, $P_{fus}(MW)$ &241&107\\
	\hline
	Heating power, $P_{{heat}}(MW)$ &74.5&3.56\\
	\hline
	\makecell{Plasma currrent, $I_p (MA)$} & 15 & 22\\
	\hline
	H factor & ($H_{98}$) 3.41&  ($H_{ST}$)0.98\\
	\hline
	$T_i/T_e$ & 4 & 4 \\
	\hline
	\makecell{p-$^{11}$B Fusion \\ enhancement factor $f_{\sigma}$} & 5 &5\\
	\hline
	\end{tabular}
	\label{tabel:CTSTcmp}
	\end{center}
\end{table}

After summarizing the key scientific issues, the key technological challenges for ST p-$^{11}$B fusion can also be outlined accordingly:
\begin{itemize}
\item High heat load materials: The compact size of ST is advantageous for the commercialization of fusion energy but results in a higher particle and energy flux on the  plasma facing components, especially the divertor. Effective management of these challenging conditions requires the development of new materials or the implementation of innovative design strategies.
\item High power supply: In addition to the power supply for the toroidal and poloidal magnetic coils, increased power demands for heating and current drive actuators are anticipated.
\item High efficient ion heating: The ion temperature required for p-$^{11}$B fusion is substantially higher than that for D-T fusion, necessitating advanced ion heating methods. Promising approaches include Neutral Beam Injection (NNBI) at high energies with long-duration operation or Ion Cyclotron Resonance Heating (ICRF). However, mature technologies for these methods are yet to be fully developed.
\item High-temperature superconductivity: Although the inherent properties of ST reduce the demand for a strong toroidal magnetic field, advancements in high-temperature superconducting materials hold the potential to optimize the engineering design of ST, making them more efficient for fusion applications.
\item Low aspect ratio engineering: The use of a low aspect ratio in fusion reactors offers benefits such as improved plasma confinement and reduced tokamak size. However, it imposes limitations on central solenoid space, affecting available volt-seconds for ohmic heating and plasma feedback control. Addressing this challenge requires the development of new technologies for effective heating and control in low aspect ratio fusion systems.
\item Advanced divertor: Controlled production of $\alpha$ particles in p-$^{11}$B fusion enables direct electric power generation. Realizing this potential requires new conceptual designs and experimental validations to ensure the feasibility and practicality of this approach.
\end{itemize}

\section{ENN fusion roadmap}\label{sec:roadmap}

In this section, we will outline a new commercial fusion roadmap for p-$^{11}$B utilizing spherical torus.

\begin{figure*}
\centering
\includegraphics[width = 0.95\textwidth]{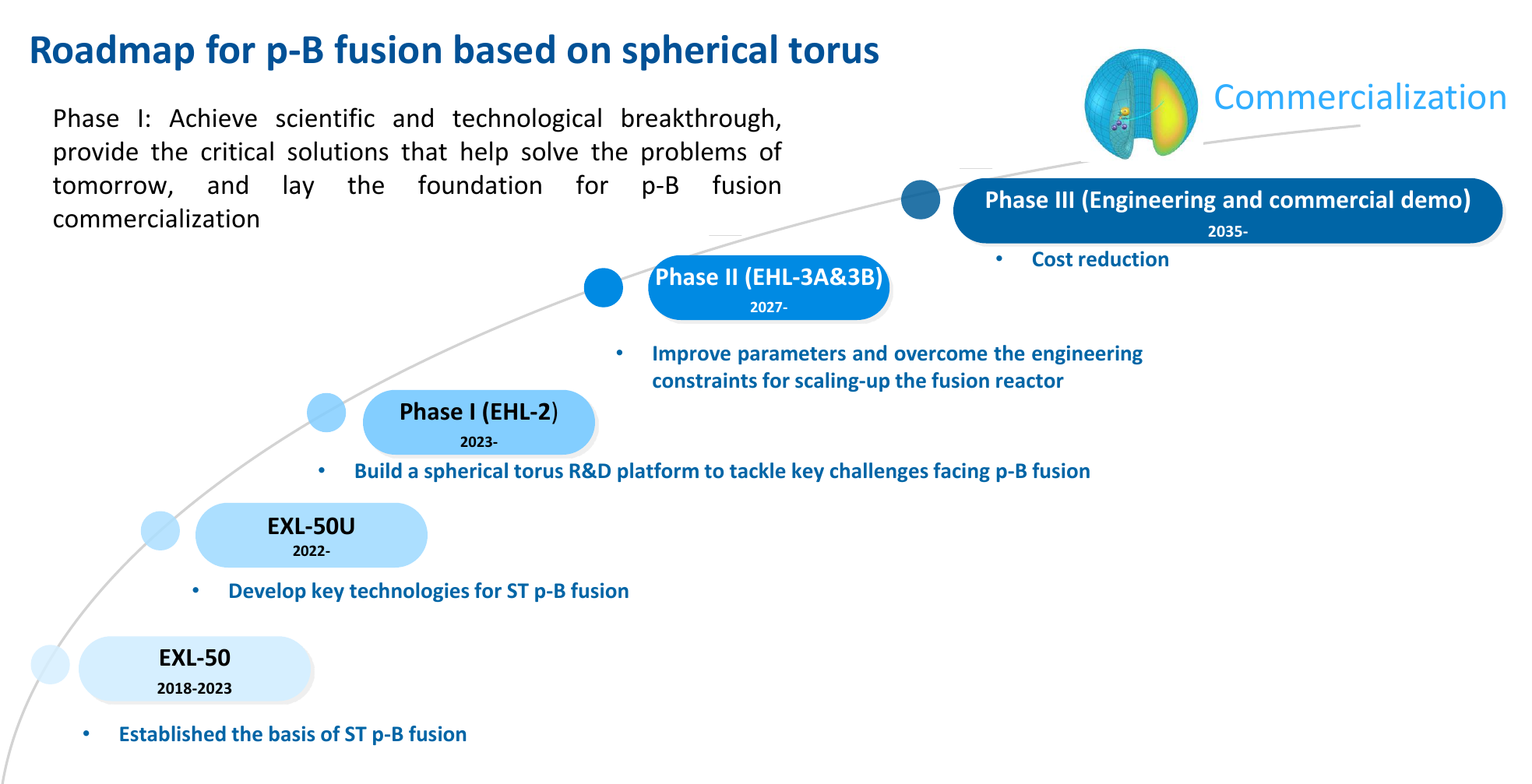}
\caption{The ENN's roadmap for ST p-$^{11}$B fusion.}\label{fig:Roadmap}
\end{figure*}

\begin{figure*}
\centering
\includegraphics[width=0.9\textwidth]{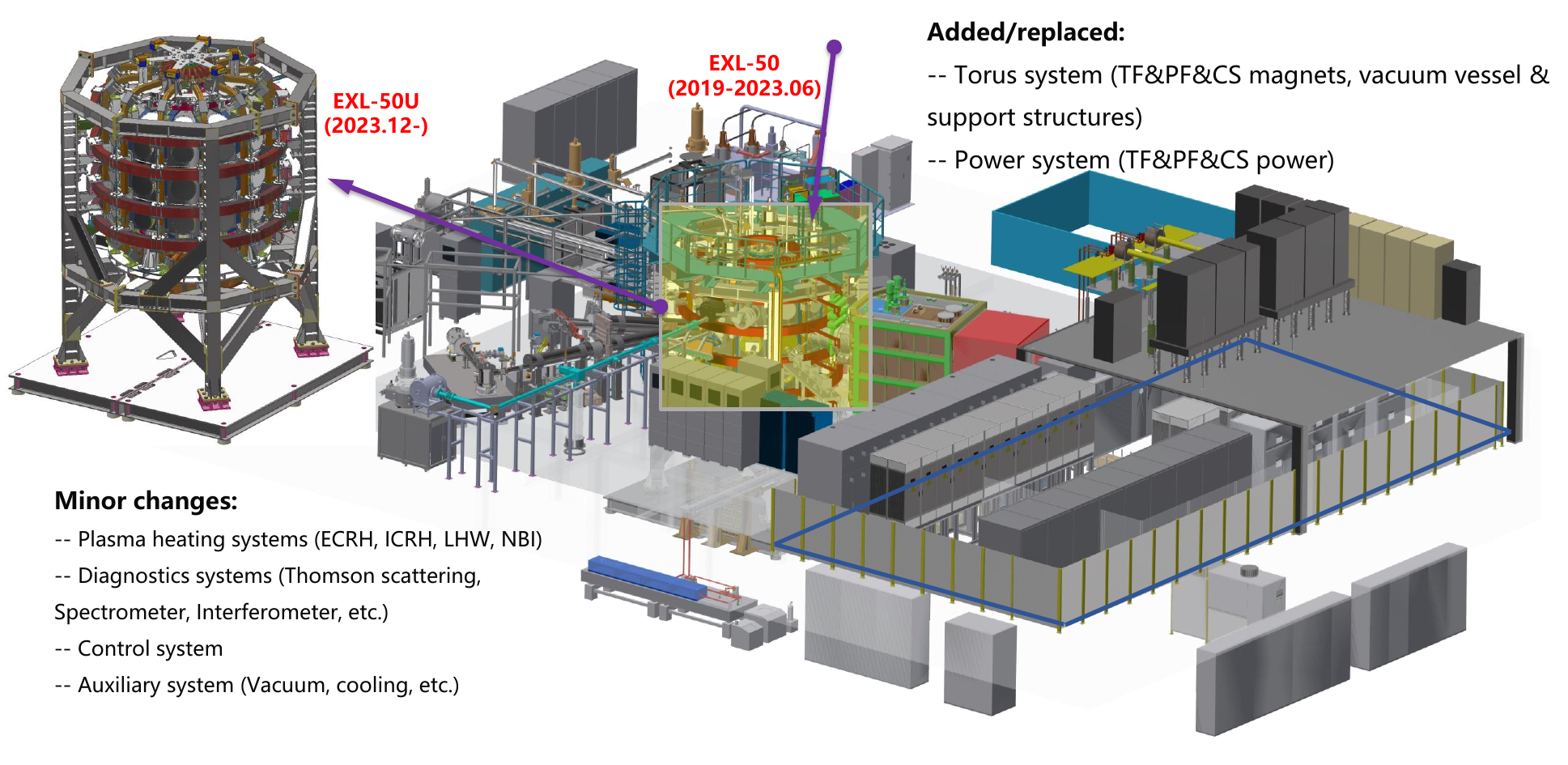}
\caption{The upgrade of EXL-50 to EXL-50U was completed in 2023. EXL-50U achieved its first plasma in January 2024.}\label{fig:EXL50U}
\end{figure*}

\begin{figure}
\centering
\includegraphics[width=0.5\textwidth]{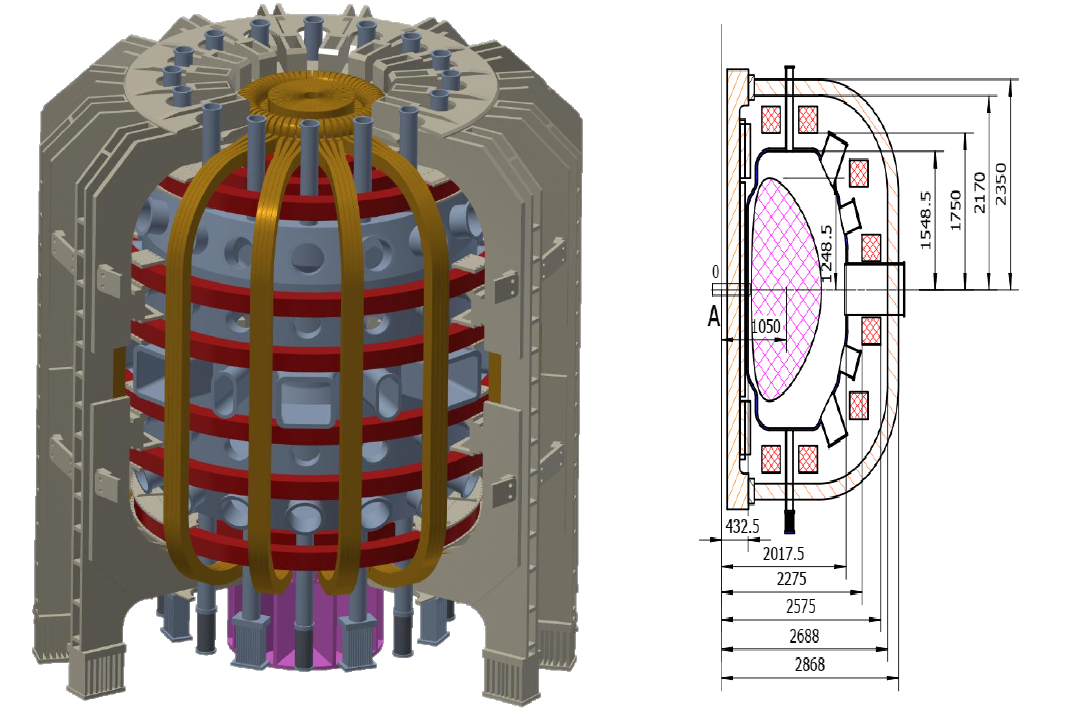}
\caption{Preliminary design of EHL-2. }\label{fig:EHL2}
\end{figure}

The roadmap for ENN fusion research is depicted in Figure \ref{fig:Roadmap}. A noteworthy milestone in this progression is the establishment of the central solenoid-free spherical torus, EXL-50, which commenced operations in 2018 (detailed in \cite{Shi2022}). EXL-50's primary focus is the exploration of current drive techniques through Electron Cyclotron Resonance Heating (ECRH). {Some initial results have been obtained, such as driving a plasma current exceeding 170 kA without a CS, and achieving a maximum driving efficiency around 1 A/W for ECRH. However, these results still need further development to achieve higher plasma density and current.} A major upgrade is on the horizon for the EXL-50 device, involving the incorporation of a central solenoid and a doubling of the toroidal magnetic field strength. This enhancement plan also encompasses improvements to NBI, ICRF, low hybrid wave heating, and ECRH. These enhancements are anticipated to facilitate the achievement of hot ion mode discharges with ion temperatures surpassing 1.0 keV. Furthermore, the subsequent project, EXL-50U (Fig.\ref{fig:EXL50U}), is designed to develop crucial technologies for ST p-$^{11}$B fusion. It also aims to conduct feasibility studies covering both the physics and engineering aspects for EHL-2 (Fig.\ref{fig:EHL2}), which is slated to be the next-generation machine in the ongoing fusion research initiative.

In the initial phase of the commercial fusion roadmap for p-$^{11}$B utilizing a spherical torus, a dedicated research and development platform, EHL-2, will be constructed. The primary objectives of Phase I include achieving the p-$^{11}$B thermal fusion reaction and validating the existence of hot ion modes characterized by a high ion-to-electron temperature ratio ($T_i/T_e \geq 2$) at elevated ion temperatures. Additionally, the platform will focus on experimental validation, specifically measuring and confirming the presence of $\alpha$ particles resulting from the fusion reaction. The research effort will explore novel physics phenomena influenced by a high magnetic field ($B_T \simeq$ 3T), conduct in-depth studies of energetic particle physics, and develop control methods for managing these energetic particles. Diagnostics related to energetic particles will be established and refined. The platform will also serve as a testing ground for a new divertor design that facilitates direct energy conversion, aiming for improved efficiency in harnessing fusion energy. Furthermore, Phase I will involve fine-tuning and optimizing strategies for achieving ion heating and current drive with maximum efficiency. Through these targeted objectives, Phase I aims to lay the groundwork for subsequent phases in the pursuit of realizing commercial p-$^{11}$B fusion using a spherical torus.

Phase II of the commercial fusion roadmap for p-$^{11}$B using a spherical torus will concentrate on elevating plasma parameters to closely emulate the conditions required for practical fusion. The construction of a new spherical torus, EXL-3, will be a key aspect of this phase, aiming to achieve a significant level of p-$^{11}$B fusion power generation. The primary objective during Phase II is to increase the ion temperature, with specific targets set at 70 keV for EHL-3A and 140 keV for EHL-3B. These targets bring the ion temperature closer to the required levels ($T_i$) for p-$^{11}$B fusion. Additionally, Phase II will involve a detailed analysis of engineering constraints that arise under fusion-related conditions, ensuring the safe and efficient operation of the fusion reactor. A critical aspect of this analysis will be the testing and selection of first-wall materials capable of withstanding the harsh nuclear environment associated with fusion reactions.

Phase III represents the apex of efforts towards achieving commercial p-$^{11}$B fusion. The primary emphasis during this phase will be on optimizing the fusion process for practical and economically viable energy generation. As part of Phase III, a comprehensive strategy for cost reduction will be explored and implemented to enhance the commercial viability of p-$^{11}$B fusion technology.

\begin{table*}
	\begin{center}
	\caption{The preliminary target parameters for ENN ST devices. {For EHL-3B, the fusion-produced alpha particles (133 MW) will also heat the plasma. In other devices listed in the table, ion heating will mainly be through NBI, with the possibility of additional heating via ICRF.}}
	\begin{tabular}{|c|c|c|c|c|c|}
	\hline
	\textbf{ Parameters} & EXL-50 & EXL-50U & EHL-2 & EHL-3A & EHL-3B \\
	\hline
	\makecell{Average / Central \\ion temperature,\\ $T_{i}$(keV)} & -/1.0 & -/6.0 &-/30 & -/70 & 46.2/140\\
	\hline
	\makecell{Average / Central \\electron density,\\ $n_{e} (10^{20}m^{-3})$} & -/0.2  &-/1.0 &-/1.3 & -/1.5 &0.83/2.5\\
	\hline
	\makecell{Confinement time,\\ $\tau_E(s)$ }& - & 0.07 & 0.5 & 5 & 25\\
	\hline
	Beta, $\beta$ & - & 0.1 & 0.11 &0.18 & 0.24\\
	\hline
	\makecell{Central Magnetic\\ field, $B_0(T)$} &0.5 & 1.2@0.6m & 3 & 4 & 4\\
	\hline
	\makecell{Major radius, \\$R_0(m)$} & 0.5 & 0.6-0.8 & 1.05 & 2 & 3.2\\
	\hline
	Aspect ratio, $A$ &1.5 & 1.5-1.85 & 1.85 & 1.8 & 1.7\\
	\hline
	\makecell{Heating power, \\$P_{heat}(MW)$} &2 & $>$3 &17&60& 6.83+133 \\
	\hline
	\makecell{Plasma currrent,\\ $I_p (MA)$} & 0.5 & 0.9 & 3.0 & 10 &25\\
	\hline
	$T_i/T_e$ & - & 1.5 & 3 & 3 & 4 \\
	\hline
	\end{tabular}
	\label{tabel:ENNST}
	\end{center}
\end{table*}

The preliminary design parameters for each device during each phase can be found in Table \ref{tabel:ENNST}.  The realization of commercial p-$^{11}$B fusion is anticipated to begin around 2035 after the successful construction and operation of these devices. { The goal of EHL-2 is to verify the thermal reaction rate of ST p-B fusion, establish experimental scaling laws, and provide a design basis for future reactors. The construction of EHL-2 is also intended to verify whether the ST scaling law still holds for high parameters. Details of the EHL-2 physics and engineering designs will be published in a series of papers over the next 1-2 years.} The construction cost of EHL-2 is estimated to be around {\textbf{5}} billion CNY, with the aim of obtaining the first plasma around 2027. 

 ENN adopts a rapid iterative approach to project development, demonstrating the capability to swiftly construct devices. For instance, EXL-50, a medium-sized ST, was built in less than 10 months (Oct. 2018 - July 2019). The upgraded device, EXL-50U, was completed in just 6 months (July 2023 - Dec. 2023). The next significant milestone and potential alternatives in the roadmap will depend on the experimental outcomes of EHL-2, expected around the year 2028.

For ion heating, ENN is primarily developing NNBI technology and investigating the feasibility of using ICRF in STs \cite{Ma2024}. The p-B reaction produces three $\alpha$ particles, posing a three-body problem where the energies of each output particle cannot be determined solely by momentum and energy conservation, unlike in D-T, D-D, and D-He3 fusion. This feature presents challenges in obtaining accurate p-B fusion cross-section data. Given the critical importance of precise cross-section data for our roadmap, ENN is also dedicating effort to measuring more accurate p-B cross-section data, with results to be published in the near future.

{  It is worth mentioning that the rapid advancement of artificial intelligence (AI) capabilities is reshaping scientific and technological research, undoubtedly accelerating progress in fusion energy research \cite{Degrave2022}. ENN is actively involved in the development of digital technology, recognizing its crucial role in accelerating progress along the ST p-B fusion energy roadmap.}

\section{Summary and Discussion}\label{sec:summ}

In this paper, we present ENN's roadmap for achieving p-$^{11}$B fusion utilizing {spherical torus} technology. {The roadmap does not present all the solutions but rather identifies the technologies we need to develop.} We conduct a comprehensive comparison of the advantages and disadvantages of different fusion reactions, ultimately selecting p-$^{11}$B fusion due to its exceptional potential for commercialization. Our analysis highlights several key factors that make p-$^{11}$B fusion a promising candidate. First, the abundance of boron fuel on Earth is expected to significantly lower the cost of p-$^{11}$B fusion compared to other fusion reactions. Additionally, the absence of a breeding blanket requirement, which is necessary for D-T fusion, simplifies the design and reduces the overall size of the fusion device. Moreover, the aneutronic nature of the p-$^{11}$B fusion reaction alleviates the need for a dense screen blanket, further contributing to size reduction. This opens up the possibility of direct energy conversion or direct electric power generation, contingent upon the controlled production of only charged $\alpha$ particles.

A simplified system code has been successfully developed, enabling the evaluation of fusion reactor performance. This code takes various input parameters into account, such as plasma shape, density and temperature profiles, magnetic field strength, plasma current, and energy confinement time. In return, it provides crucial output parameters, including plasma size, average density and temperature, fusion power, radiation loss power, heating power, fusion gain, $\beta_T$ (the ratio of plasma pressure to magnetic pressure), and scaling factors. To assess plasma confinement and performance, scaling laws for energy confinement time have been applied to both conventional tokamak and {ST}. A specific comparison between CT and ST has been conducted. The results of this comparison demonstrate that ST offer certain advantages over CT. Specifically, ST requires lower magnetic field strengths and reduced heating power, making them more feasible and potentially easier to construct for achieving the desired fusion conditions.

Fusion reactivity indeed increases with higher ion temperatures, which is advantageous for achieving successful fusion reactions. However, it's crucial to note that as temperatures rise, so does the electron bremsstrahlung power loss. When the electron bremsstrahlung power loss surpasses the fusion power at the same ion temperature, it becomes essential to operate in hot ion mode. Both CT and ST face the requirement of maintaining ion temperatures higher than electron temperatures ($T_i/T_e \geq 4$) to realize p-$^{11}$B fusion. This condition is essential to achieve the necessary conditions for successful fusion energy gain. To make p-$^{11}$B fusion more attainable, methods that enhance fusion reactivity, such as maintaining a non-thermalized ion distribution, can prove highly beneficial.

The technologies needed for p-$^{11}$B fusion can be summarized after the physical requirements have been reviewed. These include high heat load materials, high-power supply, highly efficient ion heating, high-temperature superconductivity technology, low aspect ratio engineering, and advanced divertor design.

The roadmap for ENN's p-$^{11}$B fusion using ST is outlined. Two STs, namely EXL-50 (a central solenoid-free ST) and EXL-50U (an upgraded version of EXL-50), are crucial in establishing the foundation for ST p-$^{11}$B fusion. Phase I of the roadmap has been initiated, focusing on evaluating key scientific and technological aspects, including hot ion mode experiments and validating ST confinement scaling laws at high plasma parameters. A new spherical torus, EHL-2, is being constructed during this phase, with the goal of realizing the thermal p-$^{11}$B fusion reaction and establishing the physical basis for p-$^{11}$B fusion. Following the completion of Phase I, Phase II will concentrate on improving parameters and overcoming engineering constraints to scale up the fusion reactor. EHL-3, a new ST with plasma parameters closer to those required for p-$^{11}$B fusion, will be built. In Phase III, preparations for commercial fusion will include exploring methods to reduce costs.

The roadmap for ENN's pursuit of p-$^{11}$B fusion using ST is detailed and consists of several key phases. Here's a summary of the roadmap:

Phase I - Establishing the Scientific and Technological Basis:
\begin{itemize}
\item Operation of two STs, EXL-50 (a central solenoid-free ST) and EXL-50U (an upgrade of EXL-50).
\item Evaluation of the key focuses on scientific and technological aspects.
\item Conducting hot ion mode experiments to validate their feasibility.
\item Verification of ST confinement scaling laws at high plasma parameters.
\item Construction of a spherical torus, EHL-2, dedicated to realizing the p-$^{11}$B fusion reaction.
\item Establishing the physical foundation for p-$^{11}$B fusion.
\end{itemize}

Phase II - Parameter Improvement and Engineering Overcoming:
\begin{itemize}
\item Concentration on enhancing plasma parameters and addressing engineering challenges.
\item Building EHL-3 (in two steps with EHL-3A and EHL-3B), another ST with plasma parameters approaching those required for p-$^{11}$B fusion.
\item Preparing the groundwork for scaling up the fusion reactor.
\end{itemize}

Phase III - Preparing for Commercial Fusion:
\begin{itemize}
\item Focusing on cost reduction strategies and preparations for commercial fusion.
\item Exploring methods to make p-$^{11}$B fusion economically viable.
\end{itemize}

This comprehensive roadmap outlines the stepwise progression from scientific exploration to technological development and, ultimately, the pursuit of commercial p-$^{11}$B fusion, showcasing ENN's commitment to advancing fusion energy research.

In the past five years, ENN has explored various fusion approaches, considering almost all types of fusion technologies. The findings and key challenges of commercializing fusion energy have been summarized in a report\cite{Xie2023a}. The major conclusion drawn is that achieving commercial fusion energy is highly challenging, and the success of fusion energy depends on various factors, ranging from the choice of fuel to the confinement approach. As a private company, ENN's choices are further constrained. For instance, ENN prefers not to deal with neutron-related challenges. If, in the next stages of ENN's roadmap, p-B fusion proves to be impractical, the alternative choice would be D-D-He3 fusion. This fusion process involves the catalyzed fusion of deuterium-deuterium (D-D) with a small amount of $^3$He to sustain the reactor, thereby reducing the Lawson criterion requirements. Additionally, handling neutrons in this fusion process is considered to be less challenging. Hence, the research and development efforts on spherical torus technology at ENN remain relevant, as D-D-He3 fusion is being explored as a potential alternative to p-B fusion. The focus is on achieving radiation control, high beta, and good confinement in pursuit of viable and sustainable fusion energy solutions.

\begin{figure*}
\centering
\includegraphics[width=0.7\textwidth]{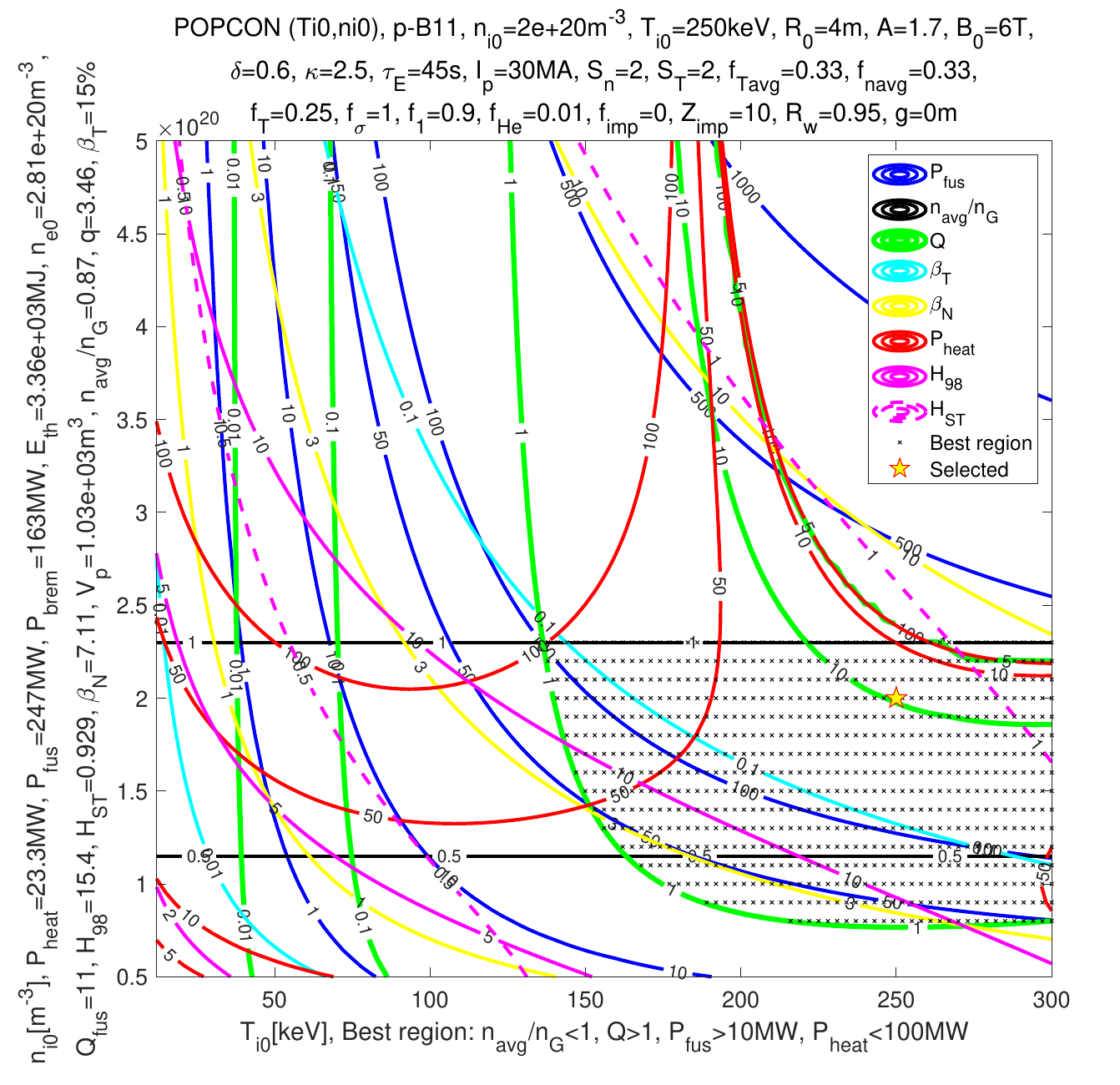}
\caption{{A typical ST p-B fusion reactor parameters, where the fusion reactivity enhancement factor $f_{\sigma}=1$, yields a fusion gain $Q_{\text{fus}}=11$. The results will change when the assumption is changed. For example, if $f_{\sigma}=2$, the required temperature $T_{i0}$ can be reduced. We provide all the important parameters in the figure for those who are interested, although this may make the figure more complicated.}}\label{fig:popconpB}
\end{figure*}

\acknowledgments
The authors express their gratitude to Mr. Yu-suo WANG, the chairman of ENN Group, for supporting this work and fusion energy research. They also acknowledge the contributions of the ENN fusion team and collaborators in supporting this endeavor. { Additionally, they would like to express their appreciation for the valuable comments and suggestions from the two anonymous referees.}

\appendix

\section{Parameter Choices for p-B Reactor}\label{sec:para}
It is straightforward to outline the rough operational parameter space for a p-B fusion reactor \cite{Xie2023a}. For instance, the optimal ion temperature falls within the range of 100-300 keV to maximize fusion power while minimizing heating requirements and radiation losses. Considering the economic and thermal loading constraints of the wall, the plasma density cannot be too high or too low, approximately $P_{\text{fus}} \approx 0.1-100 \text{MW/m}^3$, which limits $n_{e0} \approx 1-20 \times 10^{20} \text{m}^{-3}$. Furthermore, based on the Lawson criteria, the confinement time is constrained to 10-50 seconds.

For magnetic confinement fusion (MCF), since $\beta<1$, the minimum magnetic field is determined. Additionally, the magnetic field should not be excessively high to minimize cyclotron radiation losses. In the case of a ST, with $\beta\geq20\%$, the optimal magnetic field strength is approximately $B_0 \approx 5-8$ T.

To minimize bremsstrahlung and cyclotron radiation losses, the electron temperature should be kept low, ideally with $T_i/T_e\geq4$. Moreover, a high wall reflectance, preferably $\geq90\%$, is required to reduce cyclotron radiation.

Predicting the realizability of the fusion enhancement factor is challenging, although a higher value is desirable. Even for $f_{\sigma}=1$, achieving p-B fusion gain is possible \cite{Putvinski2019,Cai2022}, especially under high hot ion mode conditions ($T_i/T_e\geq4$).

In Fig.\ref{fig:popconpB}, we present a typical POPCON (Plasma Operating Contours) of a p-B ST reactor, assuming no fusion reactivity enhancement ($f_{\sigma}=1$), which still yields a fusion gain of $Q_{\text{fus}}=11$. It should be noted that different assumptions will affect the final design parameters. Therefore, the parameters provided here are for reference only. Feasibility and changes will depend on technological progress in the future. For example, if $f_{\sigma}=2$, the required temperature $T_{i0}$ can be reduced from 250 keV to 150 keV, as stated in the abstract of this work.

{

\section{More about the Reserves and Prices of Fuels}\label{sec:fuelcost}

The export and use of tritium (T) and $^3$He fuels are controlled by the government, making accurate fuel prices and reserves difficult to obtain and subject to change over time. Here, we gather some data from various sources. We assume an exchange rate of 1 USD $\approx$ 7 CNY. 
Tritium, with a half-life of 12.3 years, is reported to be priced around \$30,000 per gram (approximately CNY 0.2 million/g) \cite{Clery2022}, and commercially available tritium is estimated to be around 25 kg. 
For $^3$He, the price in 2010 was reported to be around \$2000/L (approximately CNY 0.1 million/g) \cite{Adee2010}, with an available quantity of around 10 kg.

Deuterium and boron are readily available, with the latest query prices for both being around 30 CNY/g. However, pure $^{11}$B is significantly more expensive at around 1500 CNY/g compared to natural boron powder. Global proven boron mineral mining reserves exceed one billion metric tonnes, with a natural abundance of $^{11}$B at 80.3\% \cite{WikiBoron}.

There are approximately 30 million tonnes of proven lithium reserves globally, corresponding to about 2 million tonnes of $^6$Li. The total lithium content in seawater is estimated to be 230 billion tonnes, with a natural abundance of $^6$Li at 4.85\% \cite{WikiLithium}. 

Currently, global energy consumption is approximately 20 TW \cite{WikiEnergy}. To produce 20 TW$\cdot$year of energy, equivalent to 6.3$\times$10$^{20}$ J, we can calculate the required amount of fuel using the energy yield per gram for various fusion reactions. Take the D-T reaction, for example, where 3 grams of tritium yields 1.7$\times$10$^{15}$ J of energy. Therefore, to produce 20 TW$\cdot$year of energy, 1.1$\times10^3$ kg of tritium is required. The data in Table \ref{tabel:comparison} are calculated accordingly.

}

\end{document}